\documentclass[letterpaper, 10 pt, conference]{ieeeconf}

\usepackage{graphicx}
\usepackage{epstopdf}
\usepackage{epsfig}
\usepackage{times}
\usepackage{amsmath}
\usepackage{amssymb}
\usepackage{latexsym}
\usepackage{color}
\usepackage{verbatim}
\usepackage{cite}
\usepackage{BernsteinStyle2}

\newcommand{\opq}{\textbf q}

\IEEEoverridecommandlockouts

\title{\LARGE \bf
Adaptive Target Tracking Using
Retrospective Cost Input Estimation  
}

\author{Shashank Verma$^{*}$, Sneha Sanjeevini$^{*}$,  E. Dogan Sumer$^{+}$, and Dennis S. Bernstein$^{*}$%
\thanks{$^{*}$Shashank Verma, Sneha Sanjeevini, and Dennis S. Bernstein are with the Department of Aerospace Engineering, University of Michigan, Ann Arbor, MI 48109, USA 
{\tt\small shaaero@umich.edu}}%
\thanks{$^{+}$E. S. Dogan is with Ford Motor Company.}
}

\begin{document}

\maketitle
\thispagestyle{empty}
\pagestyle{empty}




\begin{abstract}
Target tracking of surrounding vehicles is essential for collision avoidance in autonomous vehicles. Our approach to target tracking is based on causal numerical differentiation on relative position data to estimate relative velocity and acceleration.   
Causal numerical differentiation is useful for a wide range of estimation and control problems with application to robotics and autonomous systems.
The present paper extends prior work on causal numerical differentiation based on retrospective cost input estimation (RCIE).
Since the variance of the input-estimation error and its correlation with the state-estimation error (the sum of the variance and the correlation is denoted as $\widetilde{V}$)  used in the Kalman filter update are unknown, the present paper considers an adaptive discrete-time Kalman filter, where $\widetilde{V}_k$ is updated at each time step $k$ to minimize the difference between the sample variance of the innovations and the variance of the innovations given by the Kalman filter.   
The performance of this approach is shown to reach the performance of numerical differentiation based on RCIE with the best possible fixed value of $\widetilde{V}_k$. 
The proposed method thus eliminates the need to determine the best possible fixed value for $\widetilde{V}_k$.
Finally, RCIE with an adaptive Kalman filter is applied to target tracking of a vehicle using simulated data from CarSim. 
\end{abstract}



\section{INTRODUCTION}

Many applications of estimation and control benefit from the ability to perform causal differentiation, that is, numerical differentiation that provides estimates of the derivative of a signal based on current and past data \cite{Diop2000,Jauberteau2009,Stickel2010,zheng2013,reichhartinger_issueintro2018}.
One application of causal differentiation is target tracking in which velocity and acceleration are estimated from position data. Using the estimated velocity and acceleration, the future trajectory of the target can be predicted. Target tracking is important, especially in autonomous vehicles, to avoid collision with surrounding static \& non-static objects and vehicles. Target tracking based on various techniques has been considered in several works \cite{jia_2008,cui_DCN_multi_traj,florent_neural_network,song_2020,bogler_1987,kiru_2004,karsaz_2009,lee_1999,rana_2020,kalata_1984}.

Given the importance of causal numerical differentiation, it is not surprising that numerous techniques have been developed, including 
integration-based methods \cite{cullum,savitzky1964smoothing},   
observer-based methods \cite{dabroom_discrete-time_1999}, 
sliding-mode techniques \cite{arie2003slidemode,reichhartinger_arbitrary-order_2018,lopez-caamal_generalised_2019,mojallizadeh2021}, 
and input estimation \cite{shashankACC2022}.
In practice, there are three main impediments to numerical differentiation.
The first is the fact that the available signal data is sampled, and thus the idea of a ``derivative'' must be interpreted within the context of an underlying analog signal whose details are usually unknown.
Second is the fact that differentiation corresponds to an unbounded operator, which is evident by the fact that the derivative of small-amplitude high-frequency sinusoid has large amplitude.
Unboundedness implies that small-amplitude, high-frequency noise may lead to a highly inaccurate estimate of the derivative.
The challenge of numerical differentiation depends directly on the extent to which the spectrum of the signal is intertwined with the spectrum of the sensor noise.
Finally, since the derivative of the signal is unknown in practice, numerical differentiation cannot rely on an explicit error metric to determine its performance.
This impediment motivates the approach of \cite{Kutz2020}, where a Pareto-tradeoff method is used for tuning.

One approach for addressing the effect of noise is to invoke assumptions on the characteristics of the signal and the spectrum of the noise \cite{haimovich2022}.
In practice, however, it is often difficult to verify the validity of theoretical assumptions, and the effect of noise is typically addressed by employing manually tuned lowpass filtering.
%
The required filter properties are then determined manually by trial and error based on the characteristics of the sensor noise.

The present paper is motivated by the situation where the characteristics of the signal and noise are not known a priori and fixed, manual filter tuning is not effective.
This is especially the case when the spectrum of the noise is time-dependent, and changes are unpredictable, for example, due to environmental changes or sensor degradation.
In this case, there is a need for online tuning of the numerical differentiation technique to account for unknown, changing conditions.

%

By inverting the integrator dynamics, input estimation provides a technique for numerical differentiation.
In input estimation, the input to a linear system is assumed to be unknown, and the measured output of the system is used to estimate the input of the system \cite{hou1998IOIR,martin1998state,2006Floquet_state,gillijns2007unbiased,orjuela2009simultaneous,kirtikar2011delay,Fang2013HSimultaneous,yong2016unified,lu2016framework,hsieh2017unbiased,snehaZNZ,naderi2019unbiased,ZakTAC2021,ansari_input_2019}.
When the dynamics of the system consist of a cascade of one or more integrators, estimates of the input provide estimates of one or more derivatives of the output signal.
Since, like state estimation, input estimation is an online technique, this approach is suitable for causal numerical differentiation.
%
%
%

Adaptive input estimation is considered in \cite{ansari_input_2019}, where the
goal is to estimate the velocity and acceleration of a maneuvering vehicle.
The adaptive input estimator uses retrospective cost optimization to provide an estimate of the input for use by the Kalman filter.
The accuracy of this approach to causal numerical differentiation is investigated in \cite{shashankACC2022}.
%
%
%

%
%
%
%


%
Closer examination of the algorithm presented in \cite{shashankACC2022} reveals that the Kalman filter update at step $k$ depends on the sum $\widetilde{V}_k$ of the variance of the input-estimation error and its correlation with the state-estimation error.
Since the true input is unknown, it follows that $\widetilde{V}_k$ is also unknown.
The goal of the present paper is thus to extend the approach of \cite{shashankACC2022} to encompass adaptation with respect to $\widetilde{V}_k$.
In particular, the present paper adopts an adaptive Kalman filter technique that updates an estimate of $\widetilde{V}_k$.
The error metric for adaptation is given by the difference between the sample variance of the innovations and the variance of the innovations given by the Kalman filter.

Adaptive extensions of the Kalman filter to the case where the disturbance has unknown variance has been considered in \cite{Yaesh2008SimplifiedAE,shiadapukf2009,moghe2019adaptivekfLTI,Moghe2021RiemannianTB,zhangadapKF2020}. Adaptive Kalman filter based on innovations for integrating INS/GPS systems is discussed in \cite{Mohamed1999,Hide2003,Almagbile2010}. Several approaches to adaptive filtering such as bayesian, maximum likelihood, correlation, and covariance matching were studied in \cite{Mehra1972,Mehra1970}.
A related method involving a covariance constraint is developed in \cite{junkins1988minimum}.

The present paper performs causal differentiation using adaptive input estimation with an adaptive Kalman filter with the goal of applying the developed technique to target tracking.
Section II describes the retrospective cost input estimation (RCIE). Section III discusses the causal numerical differentiation problem. Section IV presents the performance metrics used in this paper. Section V describes the performance of RCIE with constant values of $\widetilde{V}_k$.Numerical examples on differentiation using RCIE with an adaptive time-varying $\widetilde{V}_k$ are presented in Section VI. Section VI also applies the proposed method to target tracking of a vehicle using simulated data from CarSim.

\section{RETROSPECTIVE COST INPUT ESTIMATION} \label{sec:InputestLinsys}

Retrospective cost input estimation (RCIE) \cite{ansari_input_2019,verma_shashank_ACC2022} can be applied to MIMO linear time-varying systems. 
The objective in this paper is to use RCIE for numerical differentiation.
For this purpose, it suffices to consider SISO linear-time invariant systems. 
Hence, the RCIE algorithm is summarized here within this context.
Consider the linear discrete-time system
\begin{align}
	x_{k+1} &=  A x_{k} + Bd_{k} + D_{1} w_{k}, 	\label{state_eqn}\\
	y_k  &= C x_k + D_{2} v_k, \label{output_eqn}
\end{align}
where
$k$ is step,
$x_k \in \mathbb R^{l_x}$ is the unknown state,
$d_k \in \mathbb R$ is the unknown input,
$w_k \in \mathbb R$ is standard Guassian white process noise,
$y_k \in \mathbb R$ is the measured output, and
$v_k \in \mathbb R$ is standard Guassian white measurement noise. 
The matrices $A \in \mathbb R^{l_x \times l_x}$, $B \in \mathbb R^{l_x \times 1}$, $D_{1} \in \mathbb R^{l_x \times 1}$, $C \in \mathbb R^{1 \times l_x}$, and $D_{2} \in \mathbb R$ are assumed to be known.
Define $V_{1} \isdef D_{1} D_{1}^\rmT$ and $V_{2} \isdef D_{2} D_{2}^\rmT$.
The goal is to estimate $d_k$ and $x_k$.

\subsection{Input Estimation}

Consider the Kalman filter forecast step
\begin{align}
	x_{{\rm fc},k+1} &= A x_{{\rm da},k} + B \hat{d}_{k},	\label{kalman_fc_state}\\
	y_{{\rm fc},k} &=  C x_{{\rm fc},k}, \label{kalman_fc_output}\\
	z_k &= y_{{\rm fc},k} - y_k, \label{innov_eq}
\end{align}
where
$\hat d_k$ is the estimate of $d_k$, 
$x_{\rm da,k} \in \mathbb R^{l_x}$ is the data-assimilation state, 
$x_{{\rm fc},k} \in \mathbb R^{l_x}$ is the forecast state, and
$z_k \in \mathbb R$ is the innovations.
The {\it input-estimation subsystem} of order $n_\rmc$ given by
\begin{align}
\hat{d}_k = \sum\limits_{i=1}^{n_\rmc} P_{i,k} \hat{d}_{k-i} + \sum\limits_{i=0}^{n_\rmc} Q_{i,k} z_{k-i}, \label{estimate_law1}
\end{align}
%
where $P_{i,k} \in \BBR$ and $Q_{i,k} \in \BBR$, is used to obtain $\hat{d}_k$.
%
%
RCIE minimizes $z_{k+1}$ by updating $P_{i,k}$ and $Q_{i,k}$.
The subsystem \eqref{estimate_law1} can be reformulated as
\begin{align}
\hat{d}_k=\Phi_k \theta_k, \label{estimate_law12}
\end{align}
where the regressor matrix $\Phi_k$ is defined by
\begin{align}
	\hspace{-0.2cm}\Phi_k \isdef
		\begin{bmatrix}
			\hat{d}_{k-1} &
			\cdots &
			\hat{d}_{k-n_{\rmc}} &
			z_k &
			\cdots &
			z_{k-n_{\rmc}}
		\end{bmatrix}
		\in \mathbb{R}^{1 \times l_{\theta}},
\end{align}
the coefficient vector $\theta_k$ is defined by
\begin{align}
\hspace{-0.2cm}\theta_k \isdef \begin{bmatrix}
P_{1,k} & \cdots & P_{n_{\rmc},k} & Q_{0,k} & \cdots & Q_{n_{\rmc},k}
\end{bmatrix}^{\rmT}
\in \mathbb{R}^{l_{\theta}},
\end{align}
and $l_\theta \isdef  2n_{\rmc} +1$.
%
%
In terms of the forward shift operator $\opq$, \eqref{estimate_law1} can be written as
\begin{align}
   \hat{d}_k = G_{\hat{d}z,k}(\opq)z_k, \label{dhat_estimate}
\end{align}
where
\begin{align}
    G_{\hat{d}z,k}(\opq) &\isdef D_{\hat{d}z, k}^{-1} (\opq) \it{N}_{\hat{d}z,k}(\opq), \label{d_hat_z_tf} \\
    D_{\hat{d}z,k}(\opq) &\isdef \opq^{n_\rmc}-P_{1,k}\opq^{n_{\rm c}-1} - \cdots-P_{n_\rmc,k}, \label{d_hat_z_tf_D} \\
    N_{\hat{d}z, k}(\opq) &\isdef Q_{0,k}\opq^{n_\rmc} + Q_{1,k} \opq^{n_\rmc-1}+\cdots+Q_{n_\rmc,k}. \label{d_hat_z_tf_N}
\end{align}
Next, define the filtered signals
%
\begin{align}
\Phi_{{\rm f},k} \isdef G_{{\rm f},k}(\opq) \Phi_k, \quad
\hat{d}_{{\rm f},k} \isdef G_{{\rm f},k}(\opq) \hat{d}_k. \label{filt_dhat}
\end{align}
%
%
Note that $G_{{\rm f},k}$ is a filter of window length $n_{\rm f} \ge 1$. 
%
Further details of the filter $G_{{\rm f}, k}$ are given in the subsection \ref{FiltConst}.
Define the {\it retrospective performance variable}
\begin{align}
z_{{\rm rc},k}(\hat{\theta}) \isdef z_k -  ( \hat{d}_{{\rm f},k} - \Phi_{{\rm f},k}\hat{\theta}   ), \label{rpv} 
\end{align}
where the coefficient vector $\hat{\theta} \in \BBR^{l_\theta}$ denotes a variable for optimization,
%
%
%
%
%
%
%
%
and define the retrospective cost function
\begin{align}
J_k(\hat{\theta}) \isdef &\sum\limits_{i=0}^k z_{{\rm rc},i}^{\rmT}(\hat{\theta}) R_z z_{{\rm rc},i}(\hat{\theta}) + \hat{\theta}^\rmT \Phi_i^\rmT R_{d} \Phi_i\hat{\theta} \nn \\
&+ (\hat{\theta} - \theta_0)^\rmT R_{\theta} (\hat{\theta} - \theta_0), \label{costf}
\end{align}
where $R_z\in\BBR$ is positive, $R_d\in\BBR$ is nonnegative, and $R_{\theta}\in\BBR^{l_{\theta} \times l_{\theta}}$ is positive definite.
Define $P_0 \isdef R_{\theta}^{-1}$. Then, for all $k\ge 0$, the cumulative cost function $J_{k}(\hat{\theta})$ has the unique global minimizer $\theta_{k+1}$ given by the RLS update
\begin{align}
P_{k+1} &= P_{k} - P_{k} \widetilde{\Phi}^{\rmT}_{k} \Gamma_{k} \widetilde{\Phi}_{k} P_{k}, \label{covariance_update} \\
\theta_{k+1} &= \theta_{k} - P_{k} \widetilde{\Phi}^{\rmT}_{k} \Gamma_{k}(\widetilde{z}_{k} + \widetilde{\Phi}_{k} \theta_{k}), \label{theta_update}
\end{align}
where
\begin{align}
\Gamma_k &\isdef ( \widetilde{R}^{-1} + \widetilde{\Phi}_k P_{k} \widetilde{\Phi}^{\rmT}_k)^{-1}, \quad
\widetilde{\Phi}_k \isdef \begin{bmatrix}
   \Phi_{\rmf, k}  \\
   \Phi_k   \\
\end{bmatrix}, \\
\widetilde{z}_k &\isdef \begin{bmatrix}
   z_k-\hat{d}_{{\rm f},k}  \\
   0   \\
\end{bmatrix}, \quad
\widetilde{R} \isdef \begin{bmatrix}
   R_z & 0  \\
   0 & R_{d}   \\
\end{bmatrix}.
\end{align}
Using the updated coefficient vector given by \eqref{theta_update}, the estimated input at step $k+1$ is given by replacing $k$ by $k+1$ in \eqref{estimate_law12}. We choose $\theta_0 = 0,$ and thus $\hat{d}_0 = 0.$
%
%

\subsection{State Estimation} \label{sec:StateEstimation}

In order to estimate the state $x_k$, $x_{{\rm fc},k}$ given by \eqref{kalman_fc_state} is used to obtain the estimate $x_{{\rm da},k}$ of $x_k$ given by the Kalman filter data-assimilation step
\begin{align}
x_{{\rm da},k} &= x_{{\rm fc},k} + K_{{\rm da},k} z_k, \label{kalman_da_state}
\end{align}
where the state estimator gain $K_{{\rm da},k} \in \mathbb R^{l_x}$ is given by
\begin{align}
	K_{{\rm da},k} &= - P_{{\rm fc},k}C^{\rmT} ( C P_{{\rm fc},k} C^{\rmT} + V_{2,k}) ^{-1}, \label{kalman_gain}
\end{align}
the data-assimilation error covariance $P_{{\rm da},k} \in \mathbb R^{l_x \times l_x}$
and the forecast error covariance $P_{{\rm fc},k} \in \mathbb R^{l_x \times l_x}$ are given by
\begin{align}
    P_{{\rm da},k} &=  (I+K_{{\rm da},k}C) P_{{\rm fc},k},\label{Pda} \\
	P_{{\rm fc},k+1} &=  A P_{{\rm da},k}A^{\rmT} + V_{1,k} + \widetilde{V}_k, \label{Pf}
\end{align} 
where $P_{{\rm fc},0} = 0$ and
\begin{align}
    \widetilde{V}_k &\isdef B {\rm var} (d_k-\hat{d}_k)B^\rmT\nn\\
    &+ A {\rm cov} (x_k - x_{{\rm da},k},d_k-\hat{d}_k)B^\rmT\nn\\
    &+ B {\rm cov} (d_k-\hat{d}_k,x_k - x_{{\rm da},k})A^\rmT.
\end{align}
Note that, since $d_k$ is unknown, it follows that $\widetilde V_k$ is unknown.
In this paper, we analyze the effect of $\widetilde V_k$ on the performance of RCIE and develop an adaptive technique for updating $\widetilde V_k$ online.

\subsection{The Filter $G_\rmf$} \label{FiltConst}
We choose $G_{{\rm f}, k}(\opq)$ to be the finite impulse response (FIR) filter
\begin{align}
G_{{\rm f}, k}(\opq) = \sum\limits_{i=1}^{n_{\rm f}} H_{i,k} \frac{1}{\textbf \opq^{i}}, \label{Gf}
\end{align}
where, for all $k\ge0$,
\begin{align}
H_{i,k} &\isdef \left\{
\begin{array}{ll}
C B, & k\ge i=1,\\
C \left(\displaystyle\prod_{j=1}^{i-1} \overline{A}_{k-j} \right) B, & k\ge i \ge 2, \\
0, & i>k,
\end{array}
\right. 
\end{align}
and $\overline{A}_k \isdef A(I + K_{{\rm da},k}C).$

\section{CAUSAL NUMERICAL DIFFERENTIATION USING RCIE}
For causal\footnote{Estimation of the derivative of the sampled $y_k$ uses the data $y_k$, and thus estimation of the derivative starts at step $k$. This implies that the estimate is not available until step $k+1$.} numerical differentiation using RCIE,
the system given by \eqref{state_eqn} and \eqref{output_eqn} is modeled as the discrete-time equivalent of an integrator. Thus, the measured output $y(t)$ is an integral of the unknown input $d(t)$ or, in other words, the unknown input $d(t)$ is the derivative of the measured output $y(t)$. Hence, by applying RCIE and reconstructing $\hat{d}$ from the estimates $\hat{d}_k$, we estimate the derivative of the measured output $y$. 
Note that process noise is not applicable when the system is modeled as an integrator. Hence, for the rest of this paper, it is assumed that $w=0,$ and thus $D_1 = 0.$

Consider the $n$-th order integrator dynamics 
\begin{align}
	\dot{x} &= A_{\rmI}x + B_{\rmI} y^{(n)}, \quad
	y = C_{\rmI}x, \label{n_int} \\
    A_{\rm I} &\isdef \begin{bmatrix}
        0_{(n-1) \times 1} & I_{n-1} \\
        0 & 0_{1\times (n-1)}
    \end{bmatrix}, \quad
    B_{\rm I} \isdef \begin{bmatrix}
        0_{(n-1)\times 1} \\ 1
    \end{bmatrix}, \label{A_hgo} \\ 
    C_{\rm I} &\isdef \begin{bmatrix}
        1 & 0_{1\times (n-1)}
    \end{bmatrix}, \label{C_hgo}
\end{align}
where $x, y \in \BBR$, and $y^{(n)}$ is the $n$-th derivative of $y$.
The discretization of \eqref{n_int} using zero-hold results in the discrete-time state space model given by
\begin{align}
    x_{k+1} &= A_{\rmd}x_k + B_{\rmd}y^{(n)}_k, \quad y_k = C_{\rmI}x_k, \label{ss_discrete} \\
    A_{\rmd} &\isdef e^{A_{\rmI}T_{\rms}}, \quad B_{\rmd} \isdef \displaystyle\int_{0}^{T_\rms}e^{A_{\rmI}(t-\tau)}B_{\rmI} d\tau, \label{AB_zh_DT}
\end{align}
where $x_k \isdef x(kT_\rms)$, $y_k \isdef y(kT_\rms)$, $y^{(n)}_k \isdef y^{(n)}(kT_\rms)$, and $T_\rms$ is the sampling time.
Setting $A = A_{\rmd}$, $B = B_{\rmd}$, and $C=C_{\rmI}$ in \eqref{state_eqn} and \eqref{output_eqn}, and applying RCIE gives an estimate $(\hat{y}^{(n)} = \hat{d})$ of $y^{(n)}$.
Note that $A_\rmd = 1, B_\rmd = T_{\rms},$ and $C_\rmI = 1$ for single differentiation, and 
\begin{align}
    A_\rmd = \begin{bmatrix}
        1 & T_\rms\\ 0 & 1
    \end{bmatrix}, \quad B_\rmd = \begin{bmatrix}
       \half T^2_\rms \\ {T_\rms}
    \end{bmatrix}, \quad C_\rmI = \begin{bmatrix}
        1 & 0
    \end{bmatrix}
\end{align}
for double differentiation.

\section{PERFORMANCE METRICS FOR RCIE}

In order to investigate the effect of $\widetilde{V}_k$ on the performance of RCIE, two metrics are defined in this section. The first metric is the normalized root-mean-square (rms) value of the input-estimation error $\hat d_k - d_k$ on $[0,k]$ given by 
\begin{align}
    \rho_k \isdef
    \sqrt{\frac{\sum_{i=0}^{k}(\hat d_i-d_i)^2}{\sum_{i=0}^{k} d_i^2}}.\label{rhodefn}
\end{align} 
Since $d$ is unknown, it follows that $\rho$ is unknown in practice, and thus it is solely used as a developmental diagnostic.
The second metric is given by 
%
%
\begin{align}
    \widetilde{S}_k \isdef |\widehat{S}_{ k}-{S}_{ k}|, \label{siev}
\end{align}
where $\widehat{S}_{ k}$ is the sample variance of $z_k$ over $[0,k]$ given by
\begin{align}
    \widehat{S}_{k} &= \cfrac{1}{k}\sum^{k}_{i=0}(z_i - \overline{z}_k)^2, \\
    \overline{z}_k &= \cfrac{1}{k+1}\sum^{k}_{i=0}z_i,
\end{align}
and ${S}_{k}$ is the value of the variance of the innovations $z_k$ given by the Kalman filter, that is,
\begin{align}
{S}_{k} \isdef  C P_{{\rm fc},k} C^{\rm T} + V_{2,k}.  
\end{align}
Note that, unlike \eqref{rhodefn}, the error metric \eqref{siev} is computable in practice.

\section{RCIE WITH CONSTANT  $\widetilde{V}_k$}\label{ConsKal}

\begin{figure}[!ht]
\vspace{-0.45cm}
\begin{center}
{\includegraphics[scale = 0.45]{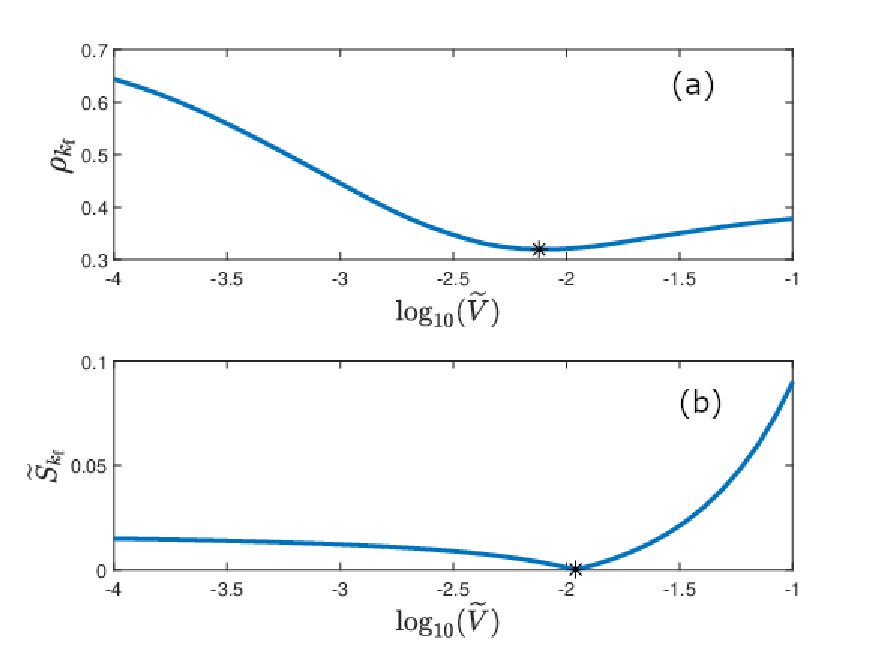}}
\vspace{-0.35cm}
\caption{Performance metrics $\widetilde{S}_{k_\rmf}$ and $\rho_{k_\rmf}$ versus the logarithm of $\widetilde{V}$ for single differentiation, where $k_\rmf = 10^4$ steps. The metrics are computed for $100$ values of $\widetilde{V}$ in the range $[10^{-6},10^2]$. However, for  clarity, the plots show values of $\widetilde{V}$ from $10^{-4}$ to $0.1$. (a) plots $\rho_{k_\rmf}$ versus $\log_{10} \widetilde{V}$. $\rho_{k_\rmf}$ is minimized at $\widetilde{V}= \widetilde{V}_{\rm sdmr} = 0.0077$. (b) plots $\widetilde{S}_{k_\rmf}$ versus $\log_{10} \widetilde{V}$. $\widetilde{S}_{k}$ is minimized at $\widetilde{V}= \widetilde{V}_{\rm sdms} = 0.0110$. } \label{fig_sd_rcie_stilde}
\end{center}
\vspace{-0.5cm}
\end{figure}

\begin{figure}[!ht]
\vspace{0cm}
\begin{center}
{\includegraphics[scale = 0.45]{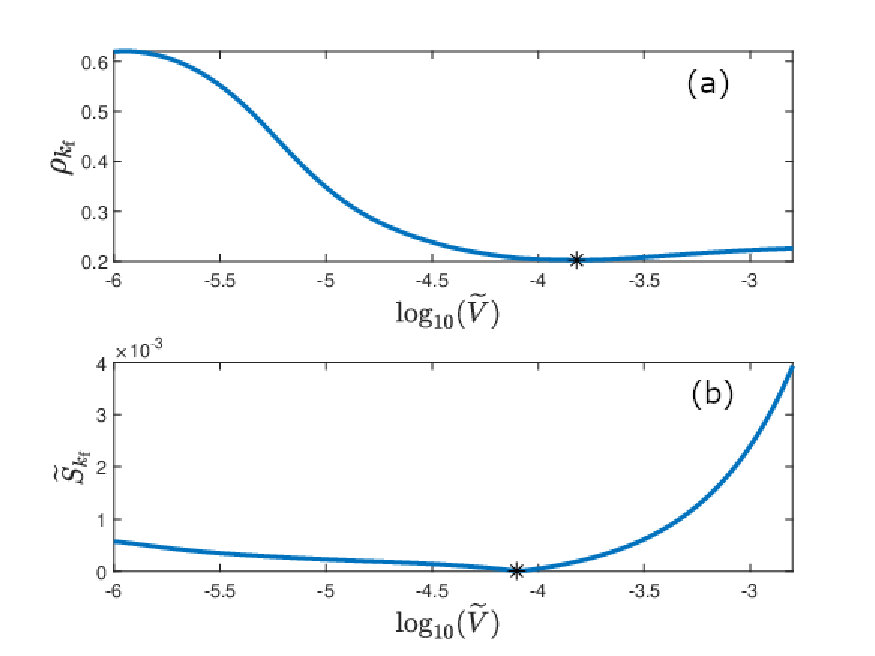}}
\vspace{-0.35cm}
\caption{Performance metrics versus the logarithm of $\widetilde{V}$ for double differentiation, where $k_\rmf = 10^4$ steps. The metrics are computed for $100$ values of $\widetilde{V}$ in the range $[10^{-6},10^{-2}]$. However, for  clarity, the plots show values of $\widetilde{V}$ from $10^{-6}$ to $10^{-2.8}$. (a) plots $\rho_{k_\rmf}$ versus $\log_{10} \widetilde{V}$. $\rho_{k_\rmf}$ is minimized at $\widetilde{V}= \widetilde{V}_{\rm sdmr} = 1.5199 \times 10^{-4}$. (b) plots $\widetilde{S}_{k_\rmf}$ versus $\log_{10} \widetilde{V}$. $\widetilde{S}_{k}$ is minimized at $\widetilde{V}= \widetilde{V}_{\rm ddms} = 7.9248 \times 10^{-5}$. } \label{fig_dd_rcie_stilde}
\end{center}
\vspace{-0.5cm}
\end{figure}

In order to assess the effect of $\widetilde{V}_k$ on $\rho_{k_\rmf}$ and $\widetilde{S}_{k_\rmf},$ we consider the case where, for all $k\ge 0$, $\widetilde{V}_k \equiv\widetilde{V}$ is a constant on $[0,k_\rmf]$ and $k_\rmf = 10000$ steps.
RCIE for both single and double differentiation is then implemented for 100 values of $\widetilde{V}$ in the range $[10^{-6},10^2]$. 
Figures \ref{fig_sd_rcie_stilde} (a) and (b) show the performance metrics $\rho_{k_{\rmf}}$ and $\widetilde{S}_{k_\rmf}$ versus $\widetilde{V}$ for single differentiation.
Similary, Figures \ref{fig_dd_rcie_stilde} (a) and (b)  show the performance metrics $\rho_{k_{\rmf}}$ and $\widetilde{S}_{k_\rmf}$ versus $\widetilde{V}$ for double differentiation. 
In the case of single differentiation, the minimum value of $\widetilde{S}_{k_\rmf}$ is achieved at $\widetilde{V}= \widetilde{V}_{\rm sdms} = 0.0110$ and the minimum value of  $\rho_{k_{\rmf}}$ is achieved at $\widetilde{V}= \widetilde{V}_{\rm sdmr} = 0.0077$.
In the case of double differentiation, the minimum value of $\widetilde{S}_{k_\rmf}$ is achieved at $\widetilde{V}= \widetilde{V}_{\rm ddms} = 7.9248 \times 10^{-5}$ and the minimum value of  $\rho_{k_{\rmf}}$ is achieved at $\widetilde{V}= \widetilde{V}_{\rm ddmr} = 1.5199 \times 10^{-4}$.

The objective of an estimation algorithm is to minimize $\rho$, however, $\rho$ is unknown in practice.
Since the values of $\widetilde{V}$ that minimize $\rho_{k_{\rmf}}$ and $\widetilde{S}_{k_\rmf}$ are close to each other and $\widetilde{S}_{k_\rmf}$ is computable in practice, $\widetilde{S}_{k_\rmf}$ is a useful metric for analyzing the performance of RCIE. 
In particular, in the next section, $\widetilde{V}_k$ is updated online by minimizing  $\widetilde{S}_{k_\rmf}$, and it is observed that using a time-varying adaptive $\widetilde{V}_k$ gives a $\rho_{k_{\rmf}}$ close to the lowest value possible using  a constant $\widetilde{V}_k$. Hence, using a time-varying adaptive $\widetilde{V}_k$ eliminates the need of guessing the optimal constant $\widetilde{V}_k$.

\section{RCIE WITH ADAPTIVE  $\widetilde{V}_k$}\label{AdapKal}

\begin{figure}[!ht]
\begin{center}
{\includegraphics[scale = 0.22]{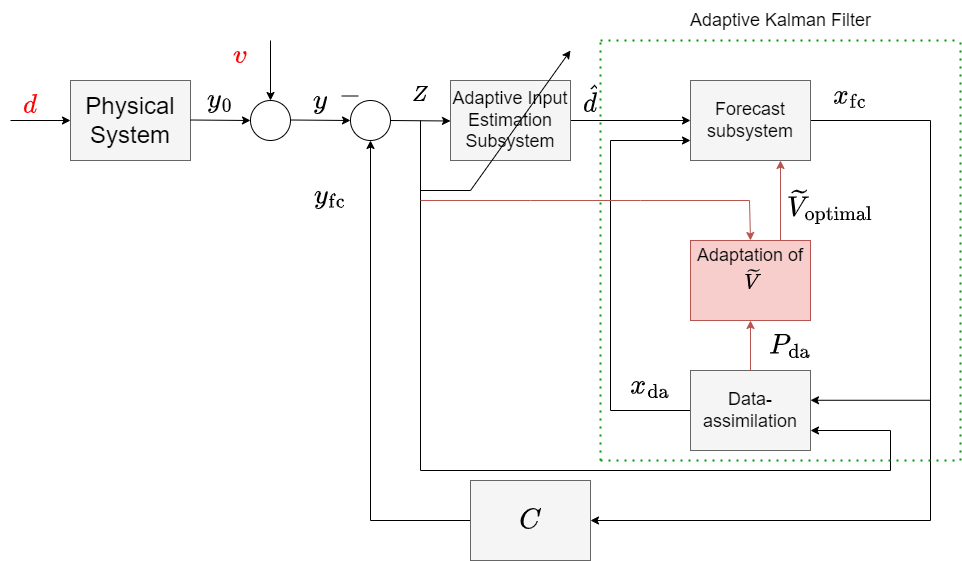}}
\vspace{-0.3cm}
\caption{{\it} Block diagram of RCIE with adaptive Kalman filter. The adaptive Kalman filter consists of the forecast subsystem, the data-assimilation subsystem, and adaption of $\widetilde{V}_k$. The innovation $z$ and the output $\hat{d}$ of the adaptive input-estimation subsystem are the inputs to the adaptive Kalman filter.} \label{fig_rcie_adtkf_block_diagram}
\end{center}
\vspace{-0.8cm}
\end{figure}

In this section, RCIE is implemented with a time-varying adaptive $\widetilde{V}_k$. In particular, $\widetilde{V}_k$ is updated at each step $k$ by choosing the value of $\widetilde{V}_k$ that minimizes $\widetilde{S}_k$.
Searching over a range of values of $\widetilde{V}_k$ by enumeration thus yields
\begin{align}
     \widetilde{V}_{{\rm opt},k}\isdef \underset{\widetilde{V}_k}{\arg \min} \ \widetilde{S}_{k}. \label{covmin}
\end{align} 
Figure \ref{fig_rcie_adtkf_block_diagram} shows the block diagram of RCIE with adaptive Kalman filter.
Numerical examples for single and double differentiation are now given to illustrate the effectiveness of RCIE with an adaptive $\widetilde{V}_k$ and compare its performance with that of RCIE with constant $\widetilde{V}_k$.
\begin{exam} \label{adapt_vtilde_sd}
{\it Adaptation of $\widetilde{V}_k$ for single differentiation.}

This example considers single differentiation in the presence of output noise using RCIE with adaptive $\widetilde{V}_k$.
Let the measured output be $y_k = \sin(0.2k)+D_2v_k$, where $v$ is standard Gaussian white noise and $D_2 = 0.0699945$.
The signal-to-noise ratio (SNR) is $20$ dB.
Let $n_\rmc = 1$, $n_\rmf = 2,$ $R_\theta = 10^{-6}I_{3}, R_d = 10^{-5}$, and $R_z = 1.$ At each step $k$, \eqref{covmin} is implemented to find $\widetilde{V}_{{\rm opt},k}$ by searching in the range $[10^{-6},10^{2}]$.

Figure \ref{fig_sd_adapt_rcie_estimate} compares the  first derivative with its estimate. 
Figure \ref{fig_sd_adapt_rcie_s_tilde_optimal} shows the evolution of $\widetilde{S}_k$ and $\widetilde{V}_{{\rm opt},k}$ versus $k$.
Figure \ref{fig_sd_adapt_const_RMSE} plots the value of $\rho_{k_\rmf}$ for RCIE with adaptive $\widetilde{V}_k$ along with the values of $\rho_{k_\rmf}$ for RCIE using the $100$ values of $\widetilde{V}$ considered in Section \ref{ConsKal}. Note that $k_\rmf = 10^4$ steps.
Figure \ref{fig_sd_rcie_rmse} plots the evolution of $\rho_k$ versus $k$ for RCIE with constant $\widetilde{V}_k = \widetilde{V}_{\rm sdmr}$ and for RCIE with adaptive $\widetilde{V}_k$.
Figures \ref{fig_sd_adapt_const_RMSE} and \ref{fig_sd_rcie_rmse} show that, at steady state, the normalized rms values of the input-estimation error for RCIE with adaptive $\widetilde{V}_k$ and RCIE with the optimal constant $\widetilde{V}_k$ are close to each other.
\begin{figure}[!ht]
\vspace{0cm}
\begin{center}
{\includegraphics[scale = 0.45]{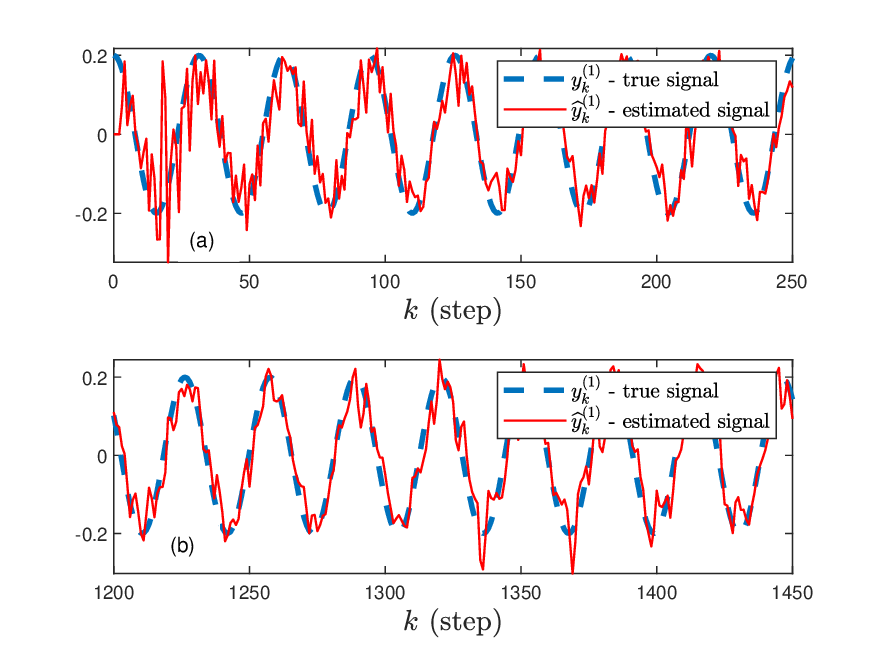}}
\vspace{-0.35cm}
\caption{{\it Example \ref{adapt_vtilde_sd}: Adaptation of $\widetilde{V}_k$ for single differentiation.} The first derivative $y^{(1)}_k$ and its estimate  $\widehat{y}^{(1)}_k$ versus $k$. (a)  $\widehat{y}^{(1)}$ follows $y^{(1)}$ after about 15 steps. (b) $\widehat{y}^{(1)}$ and $y^{(1)}$ at steady state. } \label{fig_sd_adapt_rcie_estimate}
\end{center}
\end{figure}
\begin{figure}[!ht]
\vspace{-0.4cm}
\begin{center}
{\includegraphics[scale = 0.45]{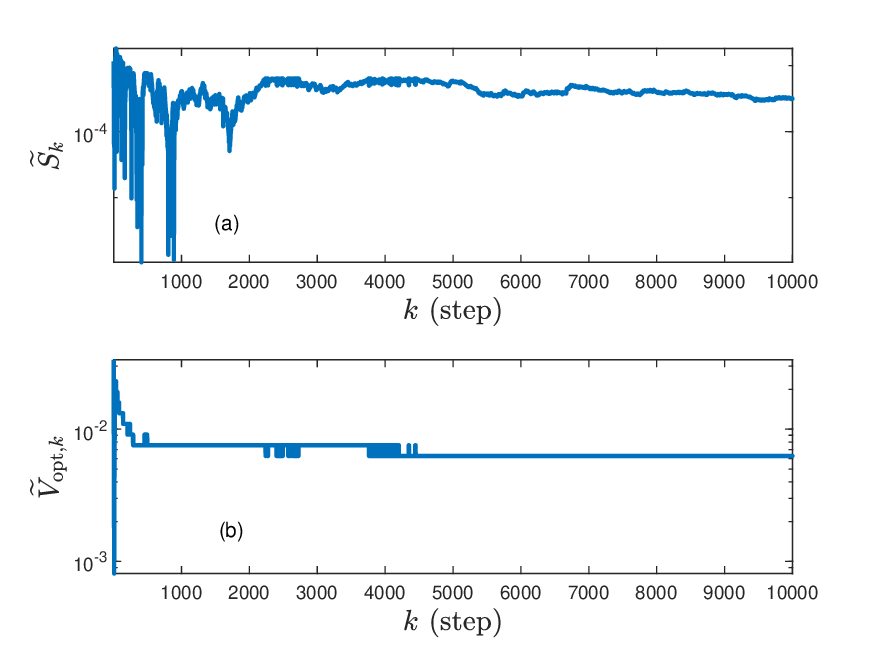}}
\vspace{-0.35cm}
\caption{{\it Example \ref{adapt_vtilde_sd}: Adaptation of $\widetilde{V}_k$ for single differentiation.} (a) plots $\widetilde{S}_k$ versus $k$. (b) plots $\widetilde{V}_{{\rm opt},k}$ versus $k$.} \label{fig_sd_adapt_rcie_s_tilde_optimal}
\end{center}
\end{figure}
\begin{figure}[!ht]
\vspace{-0.4cm}
\begin{center}
{\includegraphics[scale = 0.42]{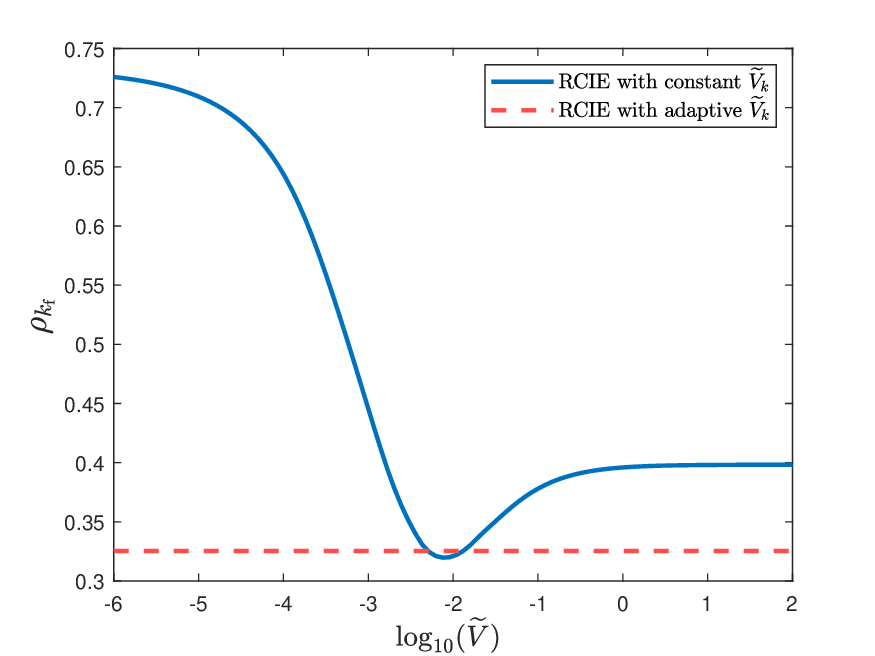}}
\vspace{-0.35cm}
\caption{{\it Example \ref{adapt_vtilde_sd}: Adaptation of $\widetilde{V}_k$ for single differentiation.} The blue solid curve is $\rho_{k_\rmf}$ versus the logarithm of  100 values of $\widetilde{V}$ in the range $[10^{-6},10^2]$. The red dashed line marks the value of $\rho_{k_\rmf}$ for RCIE with adaptive $\widetilde{V}_k$. The value of $\rho_{k_\rmf}$ for RCIE with adaptive $\widetilde{V}_k$ is close to the lowest possible value of $\rho_{k_\rmf}$ for RCIE with constant $\widetilde{V}$. Here $k_\rmf = 10^4$ steps.} \label{fig_sd_adapt_const_RMSE}
\end{center}
\vspace{-0.7cm}
\end{figure}
\begin{figure}[!ht]
\vspace{0cm}
\begin{center}
{\includegraphics[scale = 0.42]{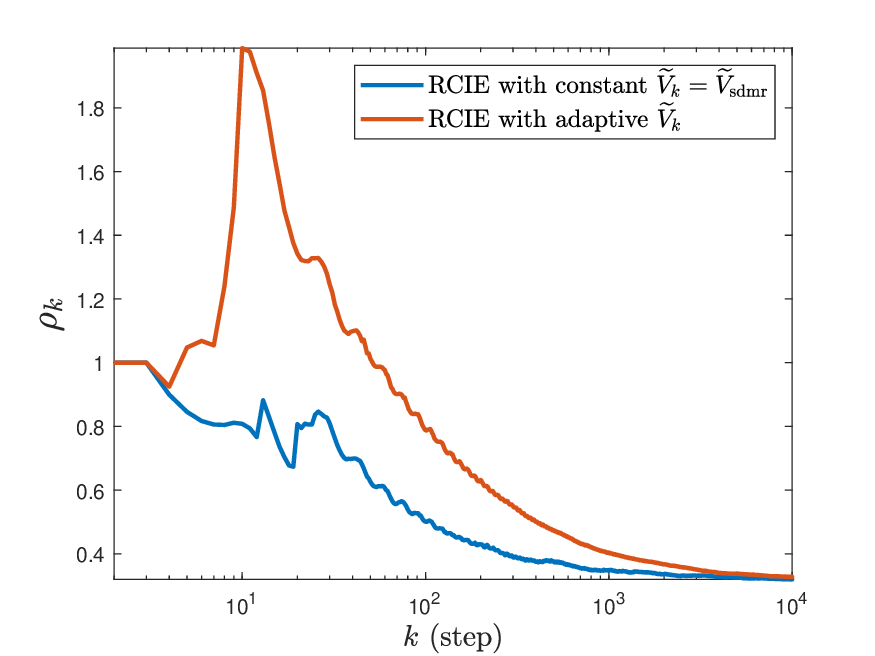}}
\vspace{-0.35cm}
\caption{{\it Example \ref{adapt_vtilde_sd}: Adaptation of $\widetilde{V}_k$ for single differentiation.} Evolution of $\rho_k$ versus $k$ for RCIE with constant $\widetilde{V}_k = \widetilde{V}_{\rm sdmr}$ and for RCIE with adaptive $\widetilde{V}_k$. At steady state, the performance of RCIE with adaptive  $\widetilde{V}_k$ is close to that of RCIE with constant $\widetilde{V}_k = \widetilde{V}_{\rm sdmr}$.} \label{fig_sd_rcie_rmse}
\end{center}
\vspace{-0.5cm}
\end{figure}
\end{exam}

\begin{exam} \label{adapt_vtilde_dd}
{\it Adaptation of $\widetilde{V}_k$ for double differentiation.}

This example considers double differentiation in the presence of output noise using RCIE with adaptive $\widetilde{V}_k$.
Let the measured output be $y_k = \sin(0.2k)+D_2v_k$, where $v$ is standard Gaussian white noise and $D_2 = 0.00699945$.
The signal-to-noise ratio (SNR) is $40$ dB.
Let $n_\rmc = 4$, $n_\rmf = 8,$ $R_\theta = 10^{-1}I_{3}, R_d = 10^{-6}$, and $R_z = 1$. At each step $k$, \eqref{covmin} is implemented to find $\widetilde{V}_{{\rm opt},k}$ by searching in the range $[10^{-6},10^{-2}]$.

Figure \ref{fig_dd_adapt_rcie_estimate} compares the second derivative with its estimate. 
Figure \ref{fig_dd_adapt_rcie_s_tilde_optimal} shows the evolution of $\widetilde{S}_k$ and $\widetilde{V}_{{\rm opt},k}$ versus $k$.
Figure \ref{fig_dd_adapt_const_RMSE} plots the value of $\rho_{k_\rmf}$ for RCIE with adaptive $\widetilde{V}_k$ along with the values of $\rho_{k_\rmf}$ for RCIE using the $100$ values of $\widetilde{V}$ considered in Section \ref{ConsKal}. Note that $k_\rmf = 10^4$ steps.
Figure \ref{fig_dd_rcie_rmse} plots the evolution of $\rho_k$ versus $k$ for RCIE with constant $\widetilde{V}_k = \widetilde{V}_{\rm ddmr}$ and for RCIE with adaptive $\widetilde{V}_k$.
Figures \ref{fig_dd_adapt_const_RMSE} and \ref{fig_dd_rcie_rmse} show that, at steady state, the normalized rms values of the input-estimation error for RCIE with adaptive $\widetilde{V}_k$ and RCIE with the optimal constant $\widetilde{V}_k$ are close to each other.
\begin{figure}[!ht]
\vspace{-0cm}
\begin{center}
{\includegraphics[scale = 0.45]{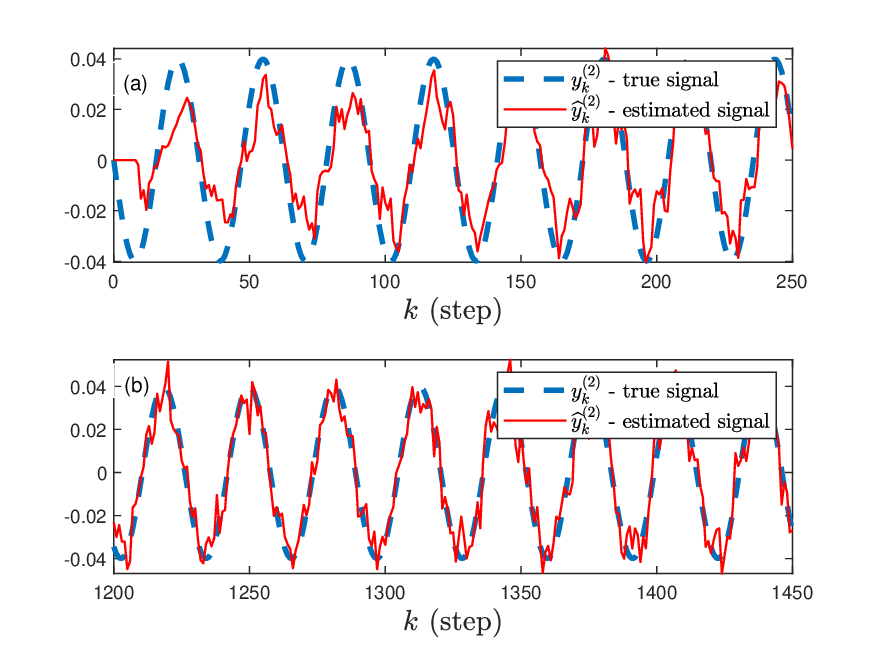}}
\vspace{-0.35cm}
\caption{{\it Example \ref{adapt_vtilde_dd}: Adaptation of $\widetilde{V}_k$ for double differentiation.} The second derivative $y_k^{(2)}$ and its 
estimate
$\widehat{y}_k^{(2)}$
versus $k$. (a)  $\widehat{y}^{(2)}$ follows $y^{(2)}$ after about 450 steps. (b) $\widehat{y}^{(2)}$ and $y^{(2)}$ at steady state. } \label{fig_dd_adapt_rcie_estimate}
\end{center}
\vspace{-0.1cm}
\end{figure}
\begin{figure}[!ht]
\vspace{-0.2cm}
\begin{center}
{\includegraphics[scale = 0.45]{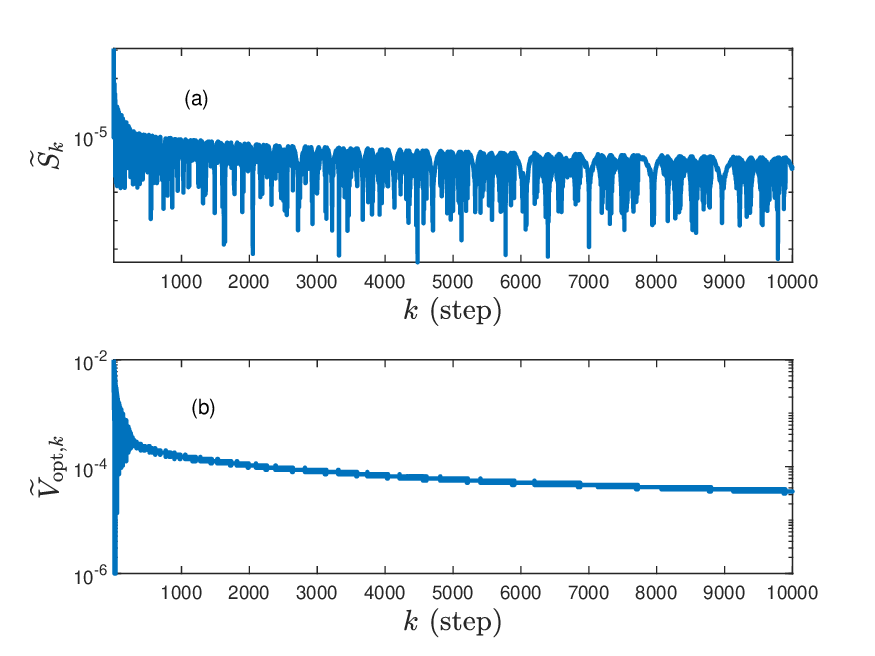}}
\vspace{-0.35cm}
\caption{{\it Example \ref{adapt_vtilde_dd}: Adaptation of $\widetilde{V}_k$ for double differentiation.} (a) plots $\widetilde{S}_k$ versus $k$. (b) plots $\widetilde{V}_{{\rm opt},k}$ versus $k$.} \label{fig_dd_adapt_rcie_s_tilde_optimal}
\vspace{-0.2cm}
\end{center}

\end{figure}
\begin{figure}[!ht]
\vspace{0cm}
\begin{center}
{\includegraphics[scale = 0.42]{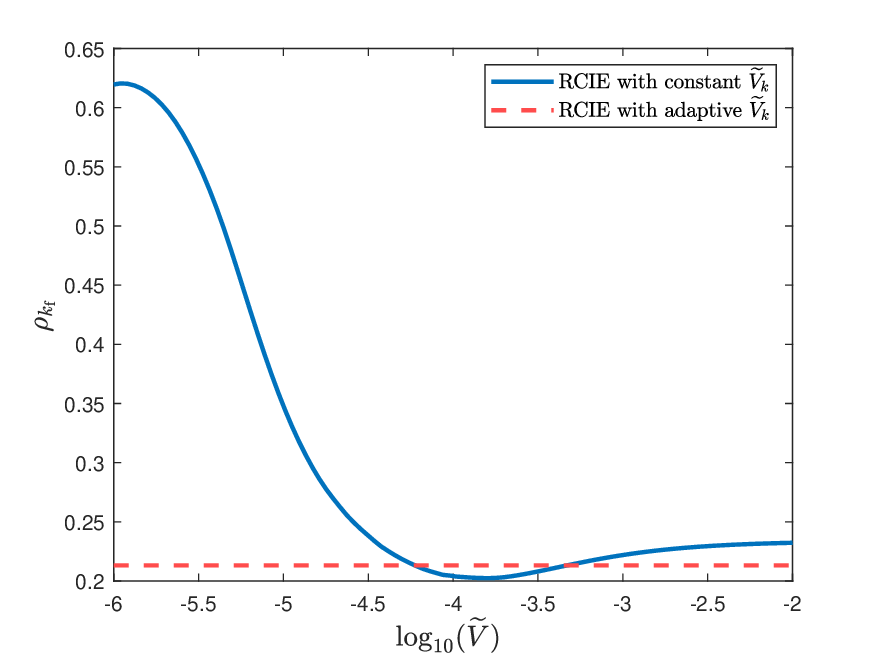}}
\vspace{-0.35cm}
\caption{{\it Example \ref{adapt_vtilde_dd}: Adaptation of $\widetilde{V}_k$ for double differentiation.} The blue solid curve is $\rho_{k_\rmf}$ versus the logarithm of  100 values of $\widetilde{V}$ in the range $[10^{-6},10^{-2}]$. The red dashed line marks the value of $\rho_{k_\rmf}$ for RCIE with adaptive $\widetilde{V}_k$. The value of $\rho_{k_\rmf}$ for RCIE with adaptive $\widetilde{V}_k$ is close to the lowest possible value of $\rho_{k_\rmf}$ for RCIE with constant $\widetilde{V}$. Here $k_\rmf = 10^4$ steps.} \label{fig_dd_adapt_const_RMSE}
\end{center}
\end{figure}
\begin{figure}[!ht]
\vspace{0cm}
\begin{center}
{\includegraphics[scale = 0.42]{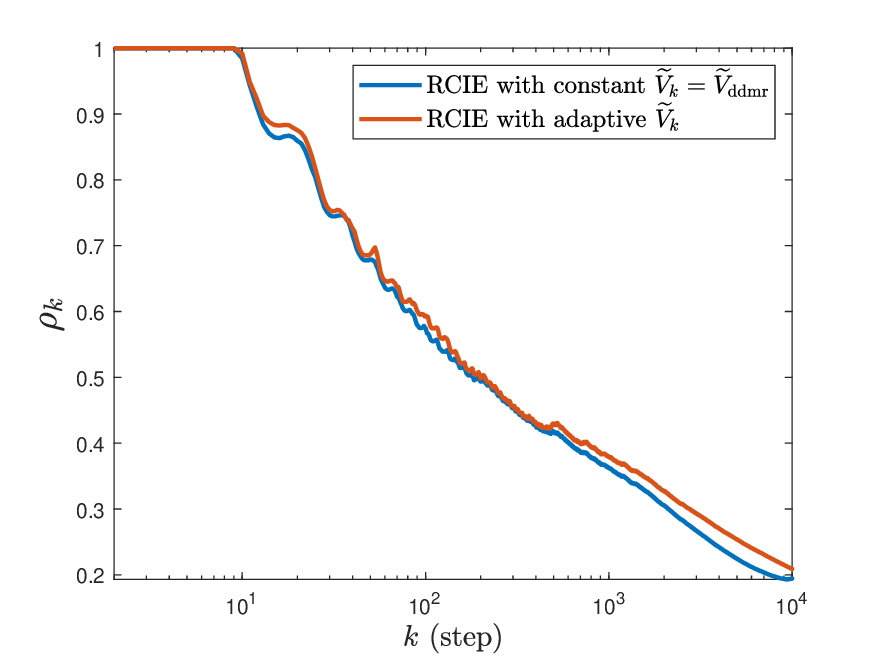}}
\vspace{-0.35cm}
\caption{{\it }{\it Example \ref{adapt_vtilde_dd}: Adaptation of $\widetilde{V}_k$ for double differentiation.} Evolution of $\rho_k$ versus $k$ for RCIE with constant $\widetilde{V}_k = \widetilde{V}_{\rm ddmr}$ and for RCIE with adaptive $\widetilde{V}_k$. The performance of RCIE with adaptive  $\widetilde{V}_k$ is close to that of RCIE with constant $\widetilde{V}_k = \widetilde{V}_{\rm ddmr}$.} \label{fig_dd_rcie_rmse}
\end{center}
\vspace{-0.5cm}
\end{figure}
\end{exam}
\begin{exam} \label{adapt_vtilde_sd_carsim}
{\it Adaptation of $\widetilde{V}_k$ for estimating the velocity of a target vehicle.}

To estimate the relative velocity of the target vehicle, RCIE with the adaptive Kalman filter is applied to simulated position data of a target vehicle relative to a host vehicle.
The CarSim simulator is used to simulate a scenario (depicted in Figure \ref{fig_carsim}) in which an oncoming target vehicle (white van) slides over to the wrong lane. The host vehicle (blue van) performs an evasive maneuver to avoid a collision. Differentiation of the relative position data along the global y-axis (shown in Figure  \ref{fig_carsim}) is done to estimate the relative velocity along the same axis. 
The same method yields and estimate of the relative velocity along the global x-axis (not shown).
%
%
The signal-to-noise ratio (SNR) of the simulated data is $40$ dB.
Let $n_\rmc = 20$, $n_\rmf = 43,$ $R_\theta = 10^{-3.2}I_{41}, R_d = 10^{-3.5}$, and $R_z = 0.98.$ 

Figure \ref{fig_sd_rcie_carsim} compares the true relative velocity with its estimate. 
Figure \ref{fig_sd_rcie_rmse_carsim} plots the value of $\rho_{k_\rmf}$ for RCIE with adaptive $\widetilde{V}_k$ along with the values of $\rho_{k_\rmf}$ for RCIE using the $200$ values of $\widetilde{V}$ considered in Section \ref{ConsKal}. Note that $k_\rmf = 2000$ steps. Figures \ref{fig_sd_rcie_carsim} and \ref{fig_sd_rcie_rmse_carsim} show that RCIE with an adaptive $\widetilde{V}_k$ gives an error that is equal to that of RCIE with the optimal constant $\widetilde{V}_k$. Hence, RCIE with adaptive Kalman filter removes the need for finding the optimal constant $\widetilde{V}_k$.
\begin{figure}[!ht]
\vspace{0.2cm}
\begin{center}
{\includegraphics[scale = 0.23]{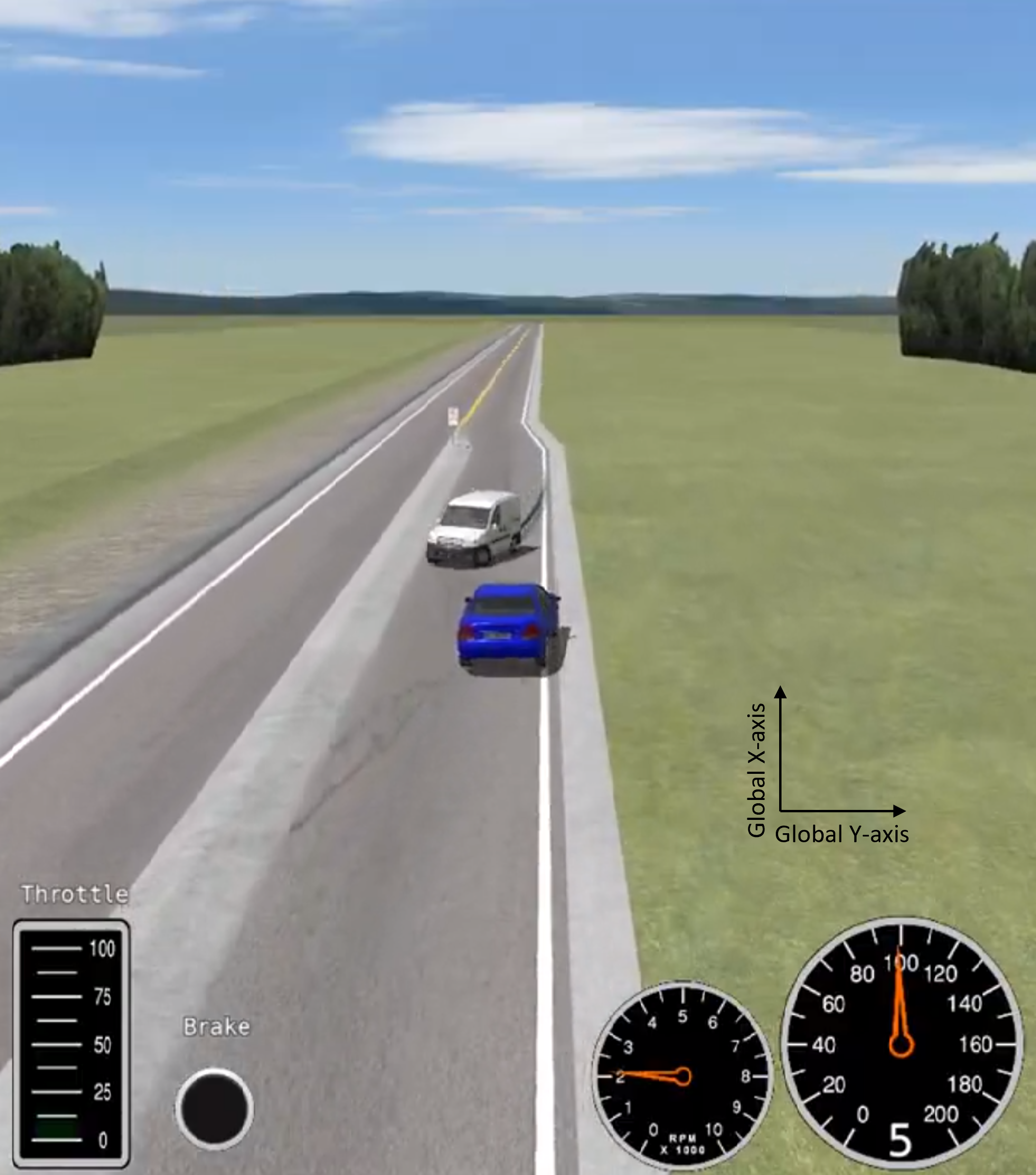}}
\vspace{-0.35cm}
\caption{{\it }{\it Example \ref{adapt_vtilde_sd_carsim}: Adaptation of $\widetilde{V}_k$ for estimation of velocity of a target vehicle.} CarSim simulated scenario in which a target vehicle (white van) slides over to the wrong lane and the host vehicle (blue van) performs evasive maneuvers to avoid a collision. } \label{fig_carsim}
\end{center}
\end{figure}
\begin{figure}[!ht]
\vspace{-0.3cm}
\begin{center}
{\includegraphics[scale = 0.42]{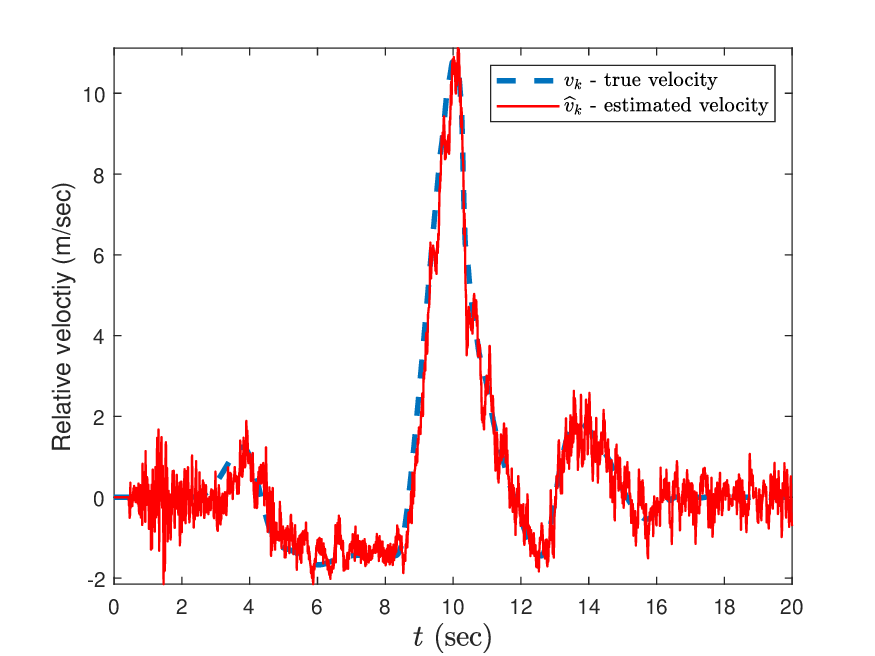}}
\vspace{-0.35cm}
\caption{{\it }{\it Example \ref{adapt_vtilde_sd_carsim}: Adaptation of $\widetilde{V}_k$ for estimation of velocity of a target vehicle.} The true relative velocity and its estimate  versus time.} \label{fig_sd_rcie_carsim}
\end{center}
\vspace{-0.3cm}
\end{figure}
\begin{figure}[!ht]
\begin{center}
{\includegraphics[scale = 0.42]{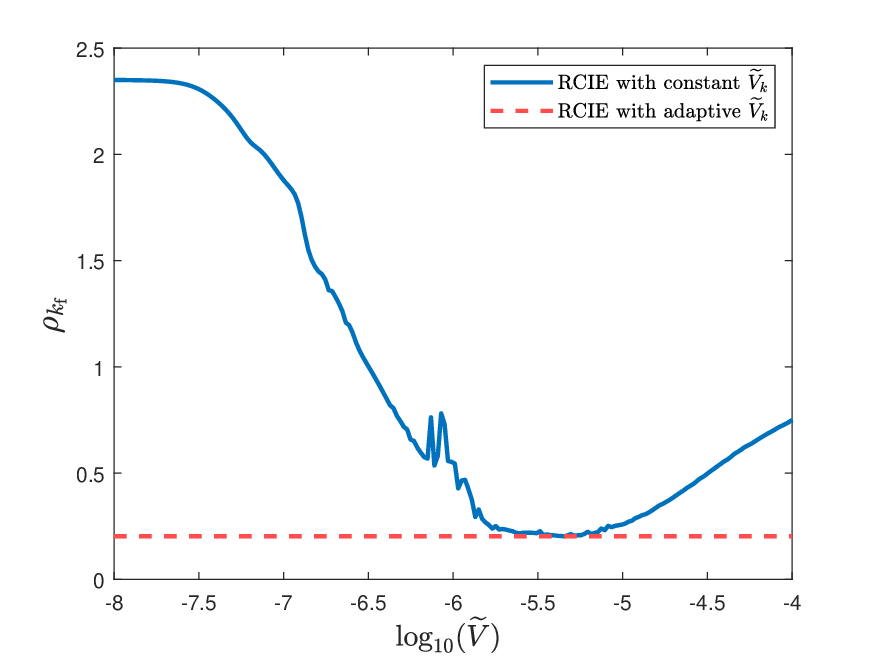}}
\vspace{-0.35cm}
\caption{{\it }{\it Example \ref{adapt_vtilde_sd_carsim}: Adaptation of $\widetilde{V}_k$ for estimation of velocity of a target vehicle.} The blue solid curve is $\rho_{k_\rmf}$ versus the logarithm of  200 values of $\widetilde{V}$ in the range $[10^{-8},10^{-4}]$. The red dashed line marks the value of $\rho_{k_\rmf}$ for RCIE with adaptive $\widetilde{V}_k$. The value of $\rho_{k_\rmf}$ for RCIE with adaptive $\widetilde{V}_k$ is close to the lowest possible value of $\rho_{k_\rmf}$ for RCIE with constant $\widetilde{V}$. Here $k_\rmf = 2000$ steps.} \label{fig_sd_rcie_rmse_carsim}
\end{center}
\vspace{-0.5cm}
\end{figure}
\end{exam}

\section{CONCLUSIONS}
This paper extended retrospective cost input estimation for causal numerical differentiation by updating the process-noise covariance $\widetilde{V}_k$ in the adaptive Kalman filter at each step.
Since $\widetilde{V}_k$ is unknown, at each time step the value that minimizes the difference between the sample variance of the  innovations and the variance of the innovations given by the Kalman filter is chosen.
It was shown that numerical differentiation based on this technique is more efficient than numerical differentiation based on a fixed value of $\widetilde{V}_k$.
RCIE with adaptive Kalman filter was tested on a collision-avoidance scenario generated by the CarSim simulator.  
Future research will extend the adaptation to include the bias and variance of the sensor noise to address the situation where the sensor noise is unknown and possibly time-dependent.
Extension to acceleration estimation will also be considered.

 \section*{ACKNOWLEDGMENTS}

This research was supported by Ford and NSF grant CMMI 2031333.
The authors are grateful to Demetrios Serakos for suggesting the use of an adaptive Kalman filter.

\bibliographystyle{ieeetr}
\bibliography{bibpaper}

\end{document}